\documentclass[a4paper,11pt]{amsart}

\usepackage[english]{my-shortcuts}
\usepackage{srcltx,algorithm,algorithmic}

\title[A Bregman Proximal ADMM for NMF with Outliers]{Estimating features with missing values and outliers: 
a Bregman-proximal point algorithm for robust Non-negative Matrix Factorization
 with application to gene expression analysis}
\author{St\'ephane Chr\'etien$^1$, Christophe Guyeux$^2$, Bastien Conesa$^3$, R\'egis Delage-Mouroux$^4$, Mich\`ele Jouvenot$^4$, \\ 
Philippe Huetz$^5$, and Fran\c coise Desc\^otes$^6$\\
\address{$^1$ Laboratoire de Mathématiques de Besançon, Universit\'e de Franche-Comt\'e,
	16, route de Gray, 25000 Besan\c{c}on, France. 
\emph{Corresponding author: stephane.chretien@univ-fcomte.fr}\\ 
$^2$~FEMTO-ST Institute, UMR 6174 CNRS, DISC Computer Science Department 
	Universit\'e de Franche-Comt\'e,
	16, route de Gray, 25000 Besan\c{c}on, France. \\
	$^3$ ISIFC, 23 Rue Alain Savary, 25000 Besançon, France. \\
	$^4$ Université de Franche-Comté, EA 3922/IFR133
Université de Franche-Comté - UFR Sciences et Techniques EA 3922/IFR133 - 25030 Besançon. \\
	$^5$ ABC\&T, 33 rue Charles Nodier, 25000 Besançon, France \\
	$^6$ Service de Biochimie et Biologie Moléculaire Sud, Pavillon 3D, Centre Hospitalier Lyon Sud, Pierre Bénite Cedex 69495, France.}}

\begin{document}


\maketitle

\begin{abstract}
To extract the relevant features in a given dataset is a difficult task, recently resolved in the non-negative data case with the 
Non-negative Matrix factorization (NMF) method.
The objective of this research work is to extend this method to the case of missing and/or corrupted data due to outliers.
To do so, data are denoised, missing values are imputed, and outliers
are detected while performing a low-rank non-negative matrix factorization of
the recovered matrix. To achieve this goal, a mixture of Bregman proximal methods and of
the Augmented Lagrangian scheme are used, in a similar way to the so-called Alternating Direction of Multipliers method. 
An application to the analysis of 
gene expression data of patients with bladder cancer is finally proposed.
\end{abstract}


%
%
%

\section{Introduction}
Finding a relevant dictionary for extracting the relevant features in a dataset is a very 
important task for many applications. The Non-negative Matrix Factorization (NMF) is a recent 
and very efficient method for achieving this goal in the case of non-negative data. Given 
a dataset consisting of $n$ vectors $x_1,\ldots,x_n$ in $\R^d$, the NMF approach 
consists of building a matrix $M$ whose columns are $x_1,\ldots,x_n$ and then factorize this matrix as
\bean 
M & = & UV^t+E, 
\eean  
where $E$ is an error term, $U$ and $V$ are componentwise non-negative, and $U$ has a small number of columns. 
The columns of $U$ represent the ``features'' present in the dataset and the interpretation of this decomposition 
is that each data consists of a mixture of the discovered features. 

Since its study by Lee and Seung \cite{LeeSeung:Nature99} in the late 90's the method, first explored in the chemometrics community, 
enjoyed a significant gain of interest from many application fields and especially in machine learning. 
It has been successfully used for document clustering \cite{XuLiuGong:SIGIR03}, email surveillance \cite{BerryMurray:CMOT05}, 
hyperspectral image analysis \cite{JiaQian:IEEEGeoRemoteSensing09}, face recognition \cite{GuillametVitria:LNAI02}, 
blind source separation \cite{Chanetal:IEEESigPro08}, etc. 
It has recently also been applied to microarray data analysis \cite{KimPark:Bioinfo07} and 
biomedicine \cite{Lietal:NMRinBiomedicine13}. 

The NMF has been computed via various approaches. One of the main strategy is the alternating minimization 
scheme, which consists in successively minimizing in $U$ and then in $V$. Since minimization 
in $U$ (resp. $V$) is a convex optimization problem, such a strategy naturally comes to mind. 
Furthermore, it has been proven very efficient in practice. However, no convergence result towards a global 
minimizer has been provided up to now. Moreover, handling the nonnegativity constraints appears 
to be delicate, since the convergence speed of the method depends on the way this constraint is 
incorporated into the iterations. A breakthrough research work concerning the possible convergence guarantees 
is~\cite{Esseretal:IEEEIP12}, where it was first proven that the problem could be convexified under
separability assumptions. Following shortly after, \cite{Bittorf:NIPS12} proposed an efficient 
approach based on linear programming also relying on separability. Separability is a property 
that can be summarized by saying that the features are some data vectors already belonging to the sample. 
Recently, under similar assumptions, 
\cite{GillisVavasis:IEEEPAMI13} proposed a very simple approach based on successive projections. 
These impressive results apply to a large set of problems where separability holds. However, separability 
does not hold in very important cases, and there is still a lot of work to do in order to 
explore the possible performance guarantees of algorithms for NMF in the future. 
Back to the not necessarily separable case, in \cite{DhillonSra:NIPS05} and \cite{Lietal:SIGKDD12}, authors proposed
Bregman divergence based iterative methods for NMF. Bregman-divergence based proximal approaches 
have been the subject of great interest recently due to 
good practical performances and  connection with mirror descent type algorithms, whose theoretical 
complexity has been extensively studied in the computational optimization community (see
for instance the survey~\cite{Bubeck:14}).  

The approach proposed in the present article relies on Bregman-proximal iterations. Our goal is to 
extend the method to the case where data may be missing and/or corrupted by the occurrence of outliers. 
Our approach borrows ideas from robust PCA \cite{Candesetal:JACM10}, where the matrix to approximate
is decomposed into a low rank part and a sparse part: \bean 
M & = & L+S.
\eean 
The low rank part $L$ is intended to approximate the data set which is supposed to be of low rank, 
and the sparse part $S$ represents the outliers. In the seminal article~\cite{Candesetal:JACM10}, the 
noise is not taken into account. However, in datasets such as gene expression data, the noise 
may be very large and one has to search for a low rank solution that removes the noise at the 
same time. This is done in the code GoDec for instance, in the case of robust PCA,
see~\cite{Godec} for further information on this method. In the present work, 
we propose an efficient method that denoises the data, guesses the missing values, and 
detects the outliers in the matrix $M$ while performing a low-rank non-negative matrix factorization of 
the recovered matrix. For this purpose, we use a mixture of Bregman proximal methods and of
the Augmented Lagrangian scheme as it is used in the Alternating Direction of Multipliers Method (ADMM). 
This mixture is also justified by the very informative recent work \cite{ParikhBoyd:FoundTrendOpt14}, 
which presents a clear interpretation of the ADMM in terms of proximal method-type iterations.  

In the next section, the Bregman proximal scheme is presented and in Section \ref{Outmiss}, 
a version taking into account potential outliers and/or missing data is described in full 
details. Then the choice of the relaxation parameters is discussed in Section \ref{lambchoice}. 
An application to the analysis of 
gene expression data of patients with bladder cancer is proposed in Section \ref{Bladder}.

\section{A Bregman proximal scheme for Non-negative Matrix Factorization}
Let $h$ be a strictly convex real valued function. Assume that $h$ is continuously differentiable
and defined on a closed convex set $\mathcal C$. Then, for all $x,y \in \mathcal C$, the Bregman 
divergence associated to $h$ is given by 
\begin{eqnarray}
D_h(y,x) & = & h(y)-h(x)-\left\langle \nabla h(x),(y-x)\right\rangle. 
\end{eqnarray}

\subsection{The space alternating Bregman-proximal scheme}
In this section, we will consider the following Bregman-proximal algorithm, which 
alternates minimization in the variable $U$ and minimization in the variable $V$:  
\begin{eqnarray}
U^{(k+1)} & \leftarrow & {\rm argmin}_{U \in \mathbb R^{d\times n}} \left\|M-UV^{(k)^{t}}\right\|_F^2+\rho D_h(U,U^{(k)}) \\
\nonumber \\
V^{(k+1)} & \leftarrow & {\rm argmin}_{V\in \mathbb R^{d\times n}} \left\|M-U^{(k+1)}V^{t}\right\|_F^2+\rho D_h(V,V^{(k)})
\end{eqnarray}
where $D_h(.,.)$ is the Bregman's divergence associated with $h(x) = x\ln(x)$, so we obtain 
\begin{eqnarray}
D_h(y,x) & = & x\ln\left(\frac{x}{y}\right)+y-x,
\end{eqnarray}
and $\rho$ is a positive constant. Let us consider the problem 		
\begin{eqnarray}
{\rm argmin}_{U\in \mathbb R^{d\times r}} \ \frac{1}{2}\left\|M-UV^{(k)^tt}\right\|_F^2+\rho \left(U^{(k)} \ln\left(\frac{U}{U^{(k)}}\right)-\left(U-U^{(k)}\right)\right).
\end{eqnarray}
The gradient of $\phi(U)$ is given by
\begin{eqnarray*}
\nabla \phi(U) & = & -(M-UV^t)V.
\end{eqnarray*}
Let us now compute the gradient of $\varphi(U)$ defined by $\varphi(U)=D_h(U,U^{(k)})$. A straightforward computation 
gives  
\begin{eqnarray*}
\frac{\partial\varphi}{\partial U_{ij}}(U) & = & \ln\left(\frac{U_{ij}}{U_{ij}^k}\right) .
\end{eqnarray*}
Therefore, taking one step in our Bregman-penalized subpace method sums up to solving  
\begin{eqnarray*}
(M-UV^t)V & = & \rho \ \ln\left(\frac{U_{ij}}{U_{ij}^k}\right).
\end{eqnarray*}
Since no explicit solution to this decoupled system of equations, we will use a fixed point approach defined as follows.

\begin{enumerate}
\item Take $U^{(k+1,0)}=U^{(k)}$.
\item $\forall \: l \in \mathbb{N}^*$, 
define 
\begin{eqnarray}
\label{fixed}
U_{ij}^{(k+1,l+1)} & = & \exp\left(\frac{1}{\rho}[(M-U^{(k+1,l)}V^t)V]_{i,j}+\ln{U_{ij}^{k}}\right) .
\end{eqnarray}
\item Stop when the difference between two successive iterates is sufficiently small, e.g., less that 1e-3. Denote 
by $l^*$ the iteration number when this occurs and output 
$U^{(k+1)}=U^{(k+1,l^*)}$.   
\end{enumerate}
The iterate $V^{(k+1)}$ can be obtained from $V^{(k)}$ using the same approach. The corresponding 
optimization problem associated to step $k+1$ is 
\begin{eqnarray*}
{\rm argmin}_{V\in \mathbb R^{n\times r}} \  \frac{1}{2}\left\|M^t-VU^{(k)^t}\right\|_F^2+\rho \left( V^{(k)} \ln\left(\frac{V}{V^{(k)}}\right)-(V-V^{(k)})\right).
\end{eqnarray*}

\subsection{A toy numerical experiment}
\begin{figure}[htb]
\begin{center}
\includegraphics[scale=0.3]{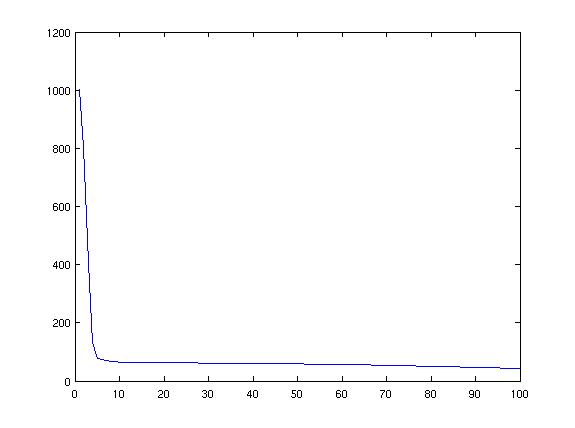}
\caption{\label{breg} Evolution of the error $M-U^{(k)}V^{(k)^t}$ as $k$ goes from 1 to 100 on 
a random example.}
\end{center}
\end{figure}

We start with a simple random example programmed in Matlab. Let $U_0$ be a random matrix in $\mathbb R^{50 \times 8}$ with i.i.d.  
components having the uniform distribution on $[0,1]$. Let $V_0$ be a random matrix in $\mathbb R^{70 \times 8}$ with 
components having the same distribution. Take $M=U_0 V_0^t$, $\rho = 100$, and random initial 
matrices. Figure \ref{breg} shows that the method converges to $M$ in the sense that it
produces a sequence of matrices $U^{(k)}$ and $V^{(k)}$ whose product $U^{(k)}V^{(k)^t}$ converges to $M$. 
\section{The case of outliers and missing data}
\label{Outmiss}
Let $\Omega$ denotes the set of couples $(i,j)$ for which an observation of $M_{i,j}$ is 
available. The matrix factorization problem can be addressed by considering the following optimization problem 
\begin{eqnarray}
\label{missout}
\min_{Y,S,U,V \geq 0}\ \frac{1}{2}\sum_{(i,j)\in \Omega} (Y_{i,j} -(UV^t)_{i,j})^2+\lambda\left\|S\right\|_1
\end{eqnarray}
subject to 
\begin{eqnarray}
M & = & Y+S ,
\end{eqnarray}  
with $\left\|S\right\|_1 = \sum_{i,j} \left|S_{ij}\right|$, and $\lambda$ is a relaxation parameter whose value is discussed in Section \ref{lambchoice}. 
\subsection{The augmented Lagrange function}
The Lagrange function $L(Y,S,U,V,\Lambda)$ for our problem is equal to:
\begin{eqnarray}
 \frac{1}{2}\sum_{(i,j)\in \Omega} (Y_{i,j} -(UV^t)_{i,j})^2
+\lambda\left\|S\right\|_1+\sum_{(i,j)\in\Omega}{\Lambda_{ij}(M_{ij}-Y_{ij}-S_{ij})} .
\end{eqnarray}
In order to enforce the constraint $M_{ij}=Y_{ij}-S_{ij}$ for all $(i,j) \in\Omega$, we introduce 
the following augmented Lagrange function 
\begin{eqnarray}
L^{aug}(Y,S,U,V,\Lambda) & = & L(Y,S,U,V)+\rho\ \frac{1}{2}\sum_{(i,j)\in\Omega}{(M_{ij}-Y_{ij}-S_{ij})^2} .
\end{eqnarray}

We now introduce an Alternating Direction Method of Multipliers. This method consists of 
solving iteratively in all the variables one after the other, and then updating the 
dual variable. For this purpose, we compute in the next subsections the optimum value for the problem of 
minimizing the augmented Lagrange function with respect to each variable. 

\subsection{Individual minimization subproblems in $Y$, $S$, $U$, and $V$}
\subsubsection{Minimization with respect to $Y$.}
The problem reduces to minimizing the function of the variable $Y$ given by 
\begin{eqnarray*}
\frac{1}{2}\sum_{i,j}({Y_{ij}-(UV^t)_{ij}})^2+\sum_{(i,j)\in\Omega}{\Lambda_{ij}(M_{ij}-Y_{ij}-S_{ij})}
+\rho\frac{1}{2}\sum_{(i,j)\in\Omega}{(M_{ij}-Y_{ij}-S_{ij})^2}
\end{eqnarray*}
Let us denote by $Y^*$ a solution to this problem. 
We have to consider two cases separately: either $(i,j) \in \Omega$, or $(i,j)\notin\Omega$. The 
case $(i,j)\notin \Omega$ is obvious, since it is straightforward to check that $Y_{ij}^*=(UV^t)_{ij}$ is 
a solution. 
Setting the partial derivative to zero gives 
the result of the case $(i,j) \in \Omega$. To summarize, we obtain
\begin{eqnarray} 
\label{Ystar}
Y_{ij}^* & = & 
\begin{cases}
\frac{1}{1+\rho}((UV^t)_{ij}+\Lambda_{ij}+\rho(M_{ij}-S_{ij})), \ \textrm{ if} (i,j) \in \Omega ,\\
\\
(UV^t)_{ij} \textrm{ otherwise}. 
\end{cases}
\end{eqnarray}
\subsubsection{Minimization with respect to $S$.}
We have to minimize the function of $S$ given by 
\begin{eqnarray}
\lambda\left\|S\right\|_1+\sum_{(i,j)\in\Omega}{\Lambda_{ij}(M_{ij}-Y_{ij}-S_{ij})}
+\rho\frac{1}{2}\sum_{(i,j)\in\Omega}{(M_{ij}-Y_{ij}-S_{ij})^2}.
\end{eqnarray}
This can be performed by optimizing each component of $S$ independently of the other. 
As for the case of minimizing with respect to $Y$, we distinguish between two cases, while if $(i,j)\notin\Omega$
one easily checks that $S_{ij}^*=0$. If $(i,j)\in \Omega$, we will use the following result. 
\begin{theo}	
The solution to \begin{eqnarray}\min_{x\in \mathbb R}\frac{1}{2}(y-x)^2+\lambda\left|x\right| 
\end{eqnarray} is given by 
\begin{eqnarray}
x^* & = &  
\begin{cases}
y-\lambda \quad if\quad  y>\lambda  ,\\
y+\lambda\quad if\quad y<\lambda  ,\\
0  \quad otherwise.\\
\end{cases} 
\end{eqnarray}
\end{theo}
Based on this result, we easily obtain 	
\begin{eqnarray}
\label{Sstar}
S_{ij}^* & = & 
\begin{cases}
-M_{ij}+Y_{ij}+\Lambda_{ij}-\frac{\lambda}{\rho}, if -M_{ij}+Y_{ij}+\Lambda_{ij}>\frac{\lambda}{\rho},\\
-M_{ij}+Y_{ij}+\Lambda_{ij}+\frac{\lambda}{\rho}, if -M_{ij}+Y_{ij}+\Lambda_{ij}<\frac{\lambda}{\rho},\\
0 \quad \textrm{ otherwise.}
\end{cases}
\end{eqnarray}

\subsubsection{Minimization with respect to $U$ and $V$.}

We just have to use the fixed point subroutine given by (\ref{fixed}).

\subsection{Our Bregman proximal-type ADMM}
We will choose the starting values as follows. Set $U^{(0)}$, $V^{(0)}$, $\Lambda^{(0)}$, $S^{(0)} = 0$. 
Set 
\begin{eqnarray*}
Y_{ij}^0 
& = & 
\begin{cases} M_{ij}, \ \forall (i,j)\in \Omega, \\
\textrm{ imputation by the mean over all other observed values row $i$}, \forall (i,j)\notin \Omega. 
\end{cases}
\end{eqnarray*}

The Bregman-Proximal point ADMM is then given by 
\begin{enumerate}
\item Set $S=S^{(k)},U=U^{(k)},V=V^{(k)},\Lambda=\Lambda^{(k)}$ and obtain $Y^{(k+1)}=Y^*$ given by (\ref{Ystar}),  
\item Set $Y=Y^{(k+1)},U=U^{(k)},V=V^{(k)},\Lambda=\Lambda^{(k)}$ and obtain $S^{(k+1)}=S^*$ given by (\ref{Sstar}), 
\item Set $Y=Y^{(k+1)},S=S^{(k+1)},U=U^{(k)},\Lambda=\lambda^{(k)}$ and obtain $U^{(k+1)}$ using the fixed point 
method (\ref{fixed}) 
\item Set $Y=Y^{(k+1)},S=S^{(k+1)},U=U^{(k+1)},\Lambda=\Lambda^{(k)}$ and obtain $U^{(k+1)}$ using the fixed point 
method (\ref{fixed}).
\item Set $\Lambda^{(k+1)}=\Lambda^{(k)}+M-Y-S$  
\end{enumerate}  

\section{Choosing the $\lambda$ value} 
\label{lambchoice}

The choice of the parameter $\lambda$ is crucial for the good performances of the proposed method. We performed 
a selection of $\lambda$ using the following approach. 

\begin{enumerate}
\item Propose an {\em a priori} range of values for $\lambda$ such that 
its maximum values leads to $S$ equals the null matrix at optimality. 
\item For each value of $\lambda$, select a set $\mathcal S$ of $s$ entries chosen uniformly at random in $M$ and consider them as missing data temporarily.
\begin{enumerate}
\item Find the solution of (\ref{missout}), where the missing data incorporate the set of entries which were artificially declared as missing in the previous step.
\item Compute the average squared error on the  data artificially declared as missing. Denote this quantity by  $err_\lambda$. 
\end{enumerate}
\item Select $\lambda$ in the prescribed range as the one which 
minimizes the average squared error $err_\lambda$. 
\end{enumerate}

\section{Application to bladder cancer expression data}
\label{Bladder}

\subsection{Description of the data}
The number of new patients affected by bladder cancer in 2013 attained 10,000 in France, thus improving the diagnosis of bladder cancer is a Public Health priority. Determining 
the genes responsible for bladder cancer would undoubtedly permit to design an efficient and adapted set of 
medical treatments. 

First of all the treatment should depend on the advancement of the tumor. For this 
purpose, researchers have gathered important gene expression data and the 
corresponding state of the malignant tumor in the bladder of 100 patients in the Lyon 
region (France). This study concerns 34 selected genes and the tumor state was discretized into 
4 classes. 

From the statistical perspective, the data can be analyzed using PCA, cluster analysis, and polytomic 
logistic regression. Our approach in this section is to use Non-negative Matrix Factorization in order
to jointly take into account the data's intrinsic non-negative nature and the necessity of clustering the data
by performing efficient feature extraction. One of the main challenges in the study of such data sets is 
to take into account possible outliers. For the bladder cancer dataset, some outliers have been 
observed by using standard PCA visualization, thus enforcing the need to automatically detect such 
phenomena in order to avoid subsequent misinterpretations of the genes' respective influences on the 
tumor state.

The data array consists of one first column providing the tumor state. 
The next 34 columns provide the expression of 34 genes. The array has 100 lines which 
correspond to the number of patients. There are two principal classes of tumors:
\begin{itemize}
\item $TVNIM$ : noninfiltrating tumors;
\item $TVIM$ : infiltrating tumors.
\end{itemize}
The tumor states have been classified into the following groups: 
\begin{itemize}
\item $Ta$ : noninfiltrating tumor in Urothelium; 
\item $T1a$ : noninfiltrating tumor in Urothelium and parts of the chorion;
\item $T1b$ : noninfiltrating tumor in Urothelium and the full chorion;
\item $>T1$ : infiltrating tumor. 
\end{itemize}
In the standard classification, the last group of the list incorporates states $T2$ to $T4b$. 

\subsection{Experimental results}

\begin{figure}
\centering
\includegraphics[scale=0.4]{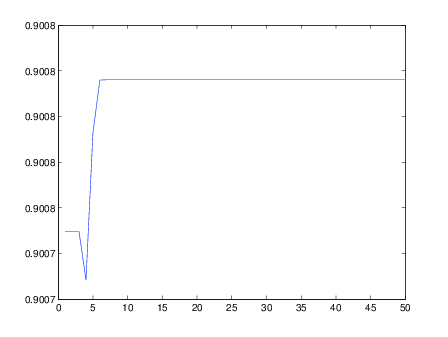}
\caption{The average squared prediction error on the artificially declared as missing entries as a function of $\lambda$}
\label{lambopt}
\end{figure}

The ADMM algorithm was run on the experimental data. The choice 
of $\lambda$ was obtained using the strategy described in 
Section~\ref{lambchoice}. 
The result obtained by this strategy is depicted in Figure \ref{lambopt}.

\begin{figure}[htbp]
\centering
\includegraphics[width=12cm]{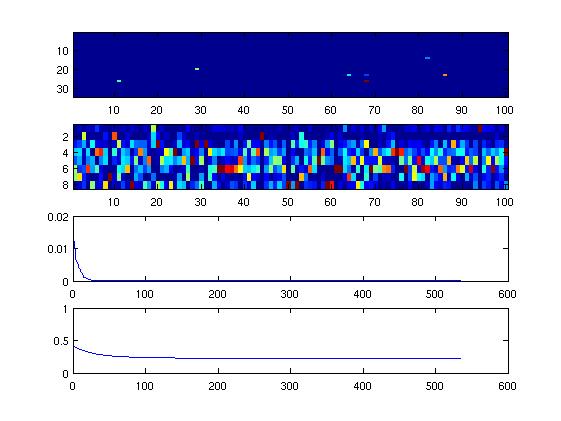}
\caption{The factorization and convergence curves. 
The first subplot is $S$ after convergence. The second subplot is $V^t$. The third subplot, shows the distance between two successive iterates of $\Lambda$. The fourth subplot shows the 
relative error between $M$ and its NMF $U^{(k)}V^{(k)^t}$ as a function 
of iteration number}
\label{ResultatsExp}
\end{figure}

\begin{figure}[htbp]
\centering
\includegraphics[width=10cm]{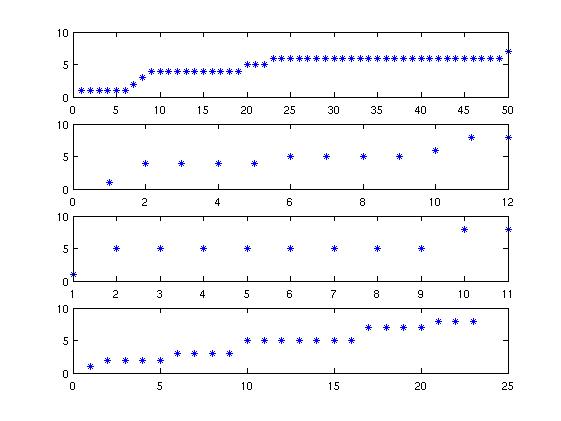}
\caption{Cluster index for each group of patients. Subplot 1 corresponds 
to pTa, subplot 2 to pT1a, subplot 3 to pT1b, and subplot 4 to > pT1.}
\label{ResultatsExp2}
\end{figure}

Based on the optimal choice of $\lambda$, the algorithm performances and the estimation results are depicted in 
Figure~\ref{ResultatsExp} and Figure~\ref{ResultatsExp2}. 

The first subplot of Figure~\ref{ResultatsExp} represents the matrix $S$ after convergence, while 
the second subplot is the matrix $V^t$. The third subplot, for its part, represents the distance in Frobenius 
norm between two successive iterates of $\Lambda$. Finally, the fourth subplot represents the evolution of the 
relative error between $M$ and its NMF $U^{(k)}V^{(k)^t}$.

Figure~\ref{ResultatsExp2} shows the cluster index in each subgroup 
"pTa", "pT1a", "pT1b", and "more serious than pT1" (\emph{i.e.}, ">pT1"). We see a sort of continuous drift in these cluster indices from pTa to >pT1. Indeed, 
\begin{itemize}
\item pTa mostly consists of 3 subgroups indexed by $\{1,4,6\}$;
\item pT1a mostly consists of 3 subgroups indexed by $\{4,5\}$;
\item pT1b mostly consists of only one group, which is $\{5\}$;
\item finally, >pT1 mostly consists of 5 subgroups indexed by $\{2,3,5,7,8\}$.
\end{itemize}
The intersection of the index subsets between two adjacent states 
is always a singleton, up to a discarded minority of individuals. 
The cluster indexed by 5 appears at medium to serious levels. The 
lowest level is characterized by cluster 6 while the most serious level 
is characterized by the more significant appearance of clusters 
2,3,7, and 8. 

Some more precise investigations should now be performed in order to 
understand the biological meaning of these clusters, \emph{i.e.}, to 
understand the factors of gravity in this cancer. 

\section{Conclusion}
In this article, a new way to find a relevant dictionary for extracting the relevant features in a given dataset has been presented, in an original context of missing values and outliers.
The well-known Non-negative Matrix Factorization (NMF) method has been extended on denoised data, where 
missing values have been guessed and outliers have been detected, leading to a mixture of Bregman proximal methods and 
the of Augmented Lagrangian scheme. Finally, an application to the analysis of 
gene expression data of patients with bladder cancer has been provided for illustration purpose.

\section{Acknowledgements} The first and second authors were funded by the grant Eléments Transposables from the Région Franche-Comté. The first author would like to acknowledge the help of Haokun Li for the preparation of the manuscript. P. Huetz
acknowledges the Montbelliard and Besan{\c c}on Leagues against Cancer for financial 
support. 

\bibliographystyle{amsplain}
\bibliography{database}

\begin{thebibliography}{10}


\bibitem{BerryMurray:CMOT05} Berry, Michael W. and Browne, Murray, Email Surveillance Using Non-negative Matrix Factorization, Computational and Mathematical Organization Theory 11, (2005), no. 3 249--264. 

\bibitem{Bittorf:NIPS12} Victor Bittorf, Benjamin Recht, C. R\'e, and Joel A. Tropp, 
Factoring non-negative matrices with linear programs, NIPS 2012.

\bibitem{Bubeck:14} S. Bubeck, Theory of convex optimization for machine learning, http://www.princeton.edu/~sbubeck/Bubeck14.pdf.

\bibitem{Candesetal:JACM10} E. J. Cand\`es, X. Li, Y. Ma, and J. Wright, 
Robust Principal Component Analysis ? Journal of ACM 58, (2010), no.1, 1-37. 

\bibitem{Chanetal:IEEESigPro08}  Chan, T.H., Ma, W.K., Chi, C.Y., Wang, Y.: A convex analysis framework for blind separation of non-negative sources. IEEE Trans. on Signal Processing, 56 (2008) no.10, 5120--5134. 

\bibitem{DhillonSra:NIPS05} Inderjit S. Dhillon, Suvrit Sra, Generalized Non-negative Matrix Approximations with Bregman Divergences, NIPS 2005.

\bibitem{Esseretal:IEEEIP12} Esser, E., Moller, M., Osher, S., Sapiro, G., Xin, J., A convex model
for non-negative matrix factorization and dimensionality reduction on physical space. IEEE Trans. on Image Processing
21 (2012) no.7, 3239--3252.

\bibitem{GillisVavasis:IEEEPAMI13} Gillis, N. and Vavasis, S., Fast and robust recursive algorithms for separable non-negative matrix factorization, IEEE Trans. Pattern Anal. Mach. Intell. (2013).
\bibitem{Godec} https://sites.google.com/site/godecomposition/home. 

\bibitem{GuillametVitria:LNAI02} Guillamet, D., Vitri\`a, J., Non-negative matrix factorization for face recognition. 
Lecture Notes in Artificial Intelligence, (2002), p. 336--344. Springer. 

\bibitem{JiaQian:IEEEGeoRemoteSensing09} Jia, S. and Qian, Y., Constrained non-negative matrix factorization for hyperspectral unmixing. IEEE Trans. on Geoscience and Remote Sensing 47, (1), p. 161--173 (2009).

\bibitem{KimPark:Bioinfo07} Hyunsoo Kim and Haesun Park, Sparse non-negative matrix factorizations via alternating non-negativity-constrained least squares for microarray data analysis, Bioinformatics 23, (2007), no. 12 1495--1502. 

\bibitem{LeeSeung:Nature99} Daniel D. Lee and H. Sebastian Seung. Learning the parts of objects by non-negative matrix factorization, Nature 401  (1999), no. 6755, p. 788--791.

\bibitem{Lietal:NMRinBiomedicine13} Li, Y., Sima, D., Van Cauter, S., Croitor Sava, A., Himmelreich, U., Pi,
Y., Van Huffel, S., Hierarchical non-negative matrix factorization (hNMF): a tissue pattern differentiation method for glioblastoma
multiforme diagnosis using MRSI. NMR in Biomedicine
26(3), 307--319 (2013).  

\bibitem{Lietal:SIGKDD12} Li, L., Lebanon, G. and Park, H., Fast Bregman divergence NMF using
Taylor expansion and coordinate descent. Proc. of the 18th ACM SIGKDD Int. Conf. on Knowledge Discovery and Data Mining, pp. 307--315 (2012).

\bibitem{ParikhBoyd:FoundTrendOpt14} N. Parikh and S. Boyd
Proximal Algorithms, Foundations and Trends in Optimization, 1(3):123-231, 2014.

\bibitem{XuLiuGong:SIGIR03} Wei Xu, Xin Liu and Yihong Gong, Document clustering based on non-negative matrix factorization, "Proceedings of the 26th annual international ACM SIGIR conference on Research and development in information retrieval". New York: Association for Computing Machinery (2003), p. 267--273.


\end{thebibliography}

\end{document}